\documentclass[aps,reprint,superscriptaddress,floatfix,nofootinbib,longbibliography]{revtex4-2}
\pdfoutput=1
\usepackage{amsmath,amssymb}
\usepackage{graphicx}
\usepackage[hidelinks]{hyperref}

\begin{document}

\title{A calibrated diagnostic for reverse-anneal sampling in programmable quantum annealers}

\author{Luis Lozano}
\email{lalozanom@tec.mx}
\affiliation{Tecnol\'ogico de Monterrey, Campus Santa Fe, Mexico City, Mexico}

\begin{abstract}
Pooling readouts from different initial preparations can conceal nearly complete initialization memory in reverse-anneal sampling.
We analyze subsystem readouts from two D-Wave processor generations using a memory order parameter and distances to a conditional-Boltzmann reference calibrated on separate single-qubit probes.
In one ten-spin condition, the probability of reading all four subsystem spins in the up state is $0.9985$, $0.0005$ and $0.1325$ for up, down and random preparations, respectively; their equal mixture is only $0.0086$ in total variation from the reference.
Across 494 Advantage2 conditions, a pooled-distance cutoff of $0.05$ selects 132 conditions, whereas additionally requiring memory at most $0.05$ selects 110.
A simpler rule requiring each preparation to lie within $0.05$ of the reference selects exactly the same 110 conditions.
Across 60 conditions on a second processor the corresponding counts are 32, 22 and 22, again with identical selections for the latter two rules.
Memory remains a target-independent diagnostic when the reference is unknown or uncertain.
Two conditions satisfying the memory cutoff, from one Ising instance at two Hamiltonian scales, strongly disagree with the prepared-boundary reference, but full-system reconstruction identifies environment reversal into the global ground state, exposing a failure of that boundary assumption.
Paired auto-scale controls change memory by approximately $0.98$ near a crossover, demonstrating the importance of submission metadata.
These observations support preparation-resolved screening and explicit validation of reference assumptions; they do not establish internal thermalization or a quantum relaxation mechanism.
\end{abstract}

\maketitle

\section{Introduction}

Programmable quantum annealers are increasingly deployed as open-system samplers for Boltzmann-machine training, probabilistic inference and hybrid optimization pipelines~\cite{adachi2015qbm,amin2018qbm,benedetti2017qbm,nelson2022gibbs,vuffray2022boltzmann}.
Any such deployment turns on two operational questions: under what conditions does a reverse-anneal protocol actually erase its initial-state memory, and what distribution does the readout represent once it has?
Initialization-dependent sampling can be useful in a finite-time heuristic, but it does not establish sampling from a common target distribution; separating these questions is therefore necessary when validating a proposed equilibrium sampler.
Existing answers are partial. A recent review poses whether a sufficiently large annealer can act as an effective bath for its own subsystems as an open research question, stating that it is not aware of such research having been performed on annealing hardware~\cite{kendon2026qacm}.
Temperature-estimation, freeze-out and Gibbs-sampling studies have shown that annealer outputs can sometimes resemble effective Boltzmann distributions while also exposing local/global and scale-dependent limitations~\cite{raymond2016global,marshall2017freezeout,marshall2019pausing,grattan2025thermometry}, and whole-device tests have compared annealer readouts against thermal targets of the pause-point Hamiltonian~\cite{izquierdo2021thermal}.
Reverse-anneal memory erasure~\cite{pelofske2025erasing}, finite-temperature criticality via quantum annealing~\cite{teza2025finite} and programmable localization~\cite{dacafilho2022localization} provide closely related but distinct context, and the theoretical foundations connecting subsystem relaxation to the structure of the global quantum state are well established~\cite{popescu2006entanglement,goldstein2006canonical,dalessio2016quantum}.
A subsystem partition makes it possible to study these questions while varying the environment size, boundary coupling, disorder and initial preparation on one programmable platform.

Here we provide such a protocol on two D-Wave QPUs with different topologies and energy scales~\cite{king2022coherent,king2023critical,mehta2025unraveling,mehta2025dwave}.
The protocol partitions the processor into a subsystem and a tunable on-chip environment, varies bath size, boundary coupling, disorder and preparation, and pairs initial-state memory with a fixed conditional-Boltzmann reference; whole-device studies of reverse-anneal memory erasure~\cite{pelofske2025erasing}, disorder-induced memory retention~\cite{dacafilho2022localization} and thermal-sampling validation~\cite{izquierdo2021thermal,nelson2022gibbs,vuffray2022boltzmann} are closely related but do not partition the processor into subsystem and bath.
We partition the annealer into a small subsystem~$S$ ($|S|=6$) coupled through programmable boundary couplers of strength $\lambda$ to an on-chip environment~$E$ (up to $|E|=50$), and use reverse annealing with a pause at $s_p=0.4$ to probe the dynamics.
We analyze the archived measurements while retaining the initial-preparation labels.  This distinction matters even when the pooled readout closely matches a target.  In a ten-spin condition with four measured subsystem spins, the probability of reading all four in the up state is $0.9985$, $0.0005$ and $0.1325$ from the up, down and random preparations, respectively; their equal mixture has total-variation distance only $0.0086$ from the calibrated reference.  Thus a pooled-distance test can select a condition whose answer depends almost completely on the initial state.

The memory order parameter $\mathcal{M}$, defined as the maximum total-variation distance between preparation-specific subsystem marginals, measures this dependence without a target distribution.  Distances to a common reference ask a different question and require that reference's assumptions to hold.  At the $0.05$ cutoffs used here, combining memory with pooled distance selects exactly the same conditions as requiring each preparation separately to match the reference: $110/494$ on Advantage2 and $22/60$ on Advantage\_system6.4.  This empirical equivalence identifies the value of preserving preparation information without claiming that the observable pair is a uniquely necessary screen.

The reference boundary also requires scrutiny.  In two large-distance conditions of one instance, full-system reconstruction identifies a global reversal into the ground state: the physical environment evolves, while the prepared-boundary reference holds it fixed.  On other memory-independent ferromagnetic conditions, nearly deterministic readouts give a small reference distance over a broad range of reference temperatures.  These comparisons distinguish initialization dependence from reference mismatch without establishing thermometry.

The absolute thresholds behind these crossovers are themselves submission-path quantities: an SDK default (\texttt{auto\_scale}) alone can shift the memory order parameter by $\Delta\mathcal{M}\approx 0.98$ on the same embedded problem in the transition region, so calibration metadata is part of the validation protocol rather than an optional log.

Throughout, the QPU is treated as the open quantum system it is~\cite{albash2018adiabatic,mehta2025dwave}: the object of study is operational subsystem relaxation and sampler validity---not isolated-system thermalization~\cite{rigol2008thermalization,dalessio2016quantum}, many-body localization~\cite{nandkishore2015mbl,abanin2019mbl}, or quantum computational advantage---and the contribution is a subsystem-level validation protocol anchored to a fixed, independently calibrated effective inverse temperature used as a discrepancy reference.

\section{Results}

\subsection{Reverse-anneal protocol and subsystem partition}

\begin{figure*}[!htbp]
\centering
\includegraphics[width=\textwidth]{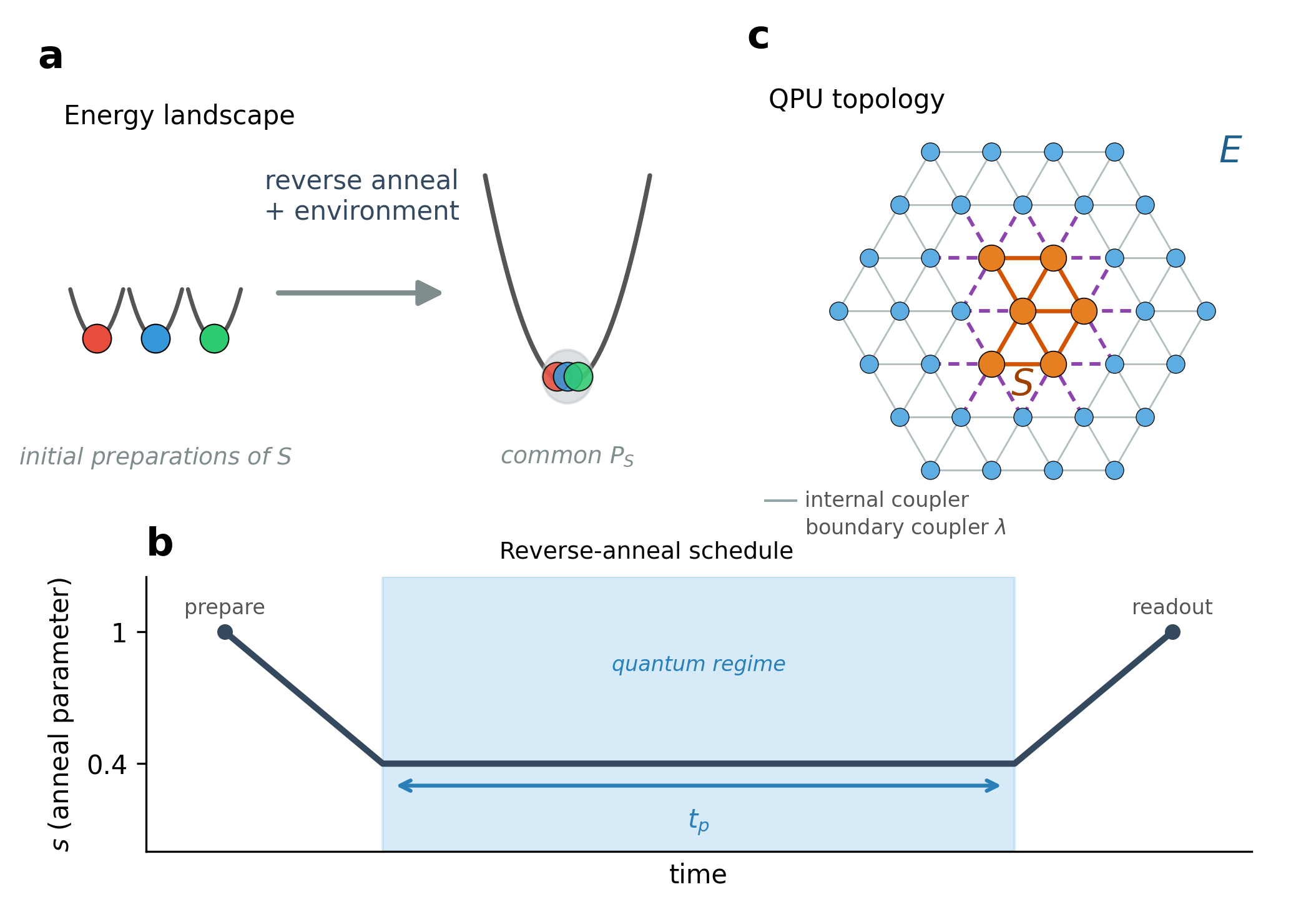}
\caption{\textbf{Reverse-anneal protocol and subsystem--environment partition.}
\textbf{a,}~Energy-landscape concept: three distinct initial preparations of the subsystem (colored balls in different potential wells) converge to the same distribution under relaxing conditions after reverse annealing in the presence of a large on-chip environment.
\textbf{b,}~Reverse-anneal schedule.  The system is initialized at $s=1$ (classical), ramped to a pause point $s_p$ where quantum fluctuations are reintroduced (blue-shaded region), held for a time $t_p$, and returned to $s=1$ for readout.
\textbf{c,}~Schematic of the subsystem--environment partition on the QPU topology.  The subsystem~$S$ (orange, $|S|=6$) is a connected cluster of qubits; the environment~$E$ (blue) surrounds it.  Dashed purple lines indicate the boundary couplers of programmable strength~$\lambda$.}
\label{fig:protocol}
\end{figure*}

Our experiment uses D-Wave's reverse-annealing protocol (Fig.~\ref{fig:protocol}b).
The QPU implements the time-dependent Hamiltonian $H(s) = -\frac{A(s)}{2}\sum_i \sigma_x^i + \frac{B(s)}{2}H_P$, where $H_P = \sum_i h_i \sigma_z^i + \sum_{ij} J_{ij}\sigma_z^i\sigma_z^j$ is the programmable problem Hamiltonian and $s\in[0,1]$ parameterizes the anneal~\cite{dwave2024reverse_anneal}.
In reverse annealing, the system is initialized in a classical state at $s=1$, ramped to a pause point~$s_p < 1$ where quantum fluctuations are reintroduced, held for a pause time~$t_p$, and returned to $s=1$ for readout.

We partition the qubits into a subsystem~$S$ ($|S|=6$) and an environment~$E$ ($|E|$ up to 50), connected through boundary couplers of strength~$\lambda$ (Fig.~\ref{fig:protocol}c).
Internal couplings are ferromagnetic ($J=-1$), and quenched disorder is introduced as random longitudinal fields $h_i\sim\mathcal{U}[-W,W]$.
Different initial preparations of~$S$ (all spins up, all down, or random) are used with the same initial preparation of~$E$.  The environment is free to evolve during the protocol.
The key observable is the memory order parameter~$\mathcal{M}$, defined as the maximum pairwise total variation distance (TVD) between the marginal distributions $P_S$ obtained from different initial states (see Methods).
$\mathcal{M}\to 0$ signals initial-state independence; $\mathcal{M}= 1$ indicates complete memory retention.

\subsection{Subsystem relaxation with tunable environment}

\begin{figure*}[!htbp]
\centering
\includegraphics[width=\textwidth,height=0.55\textheight,keepaspectratio]{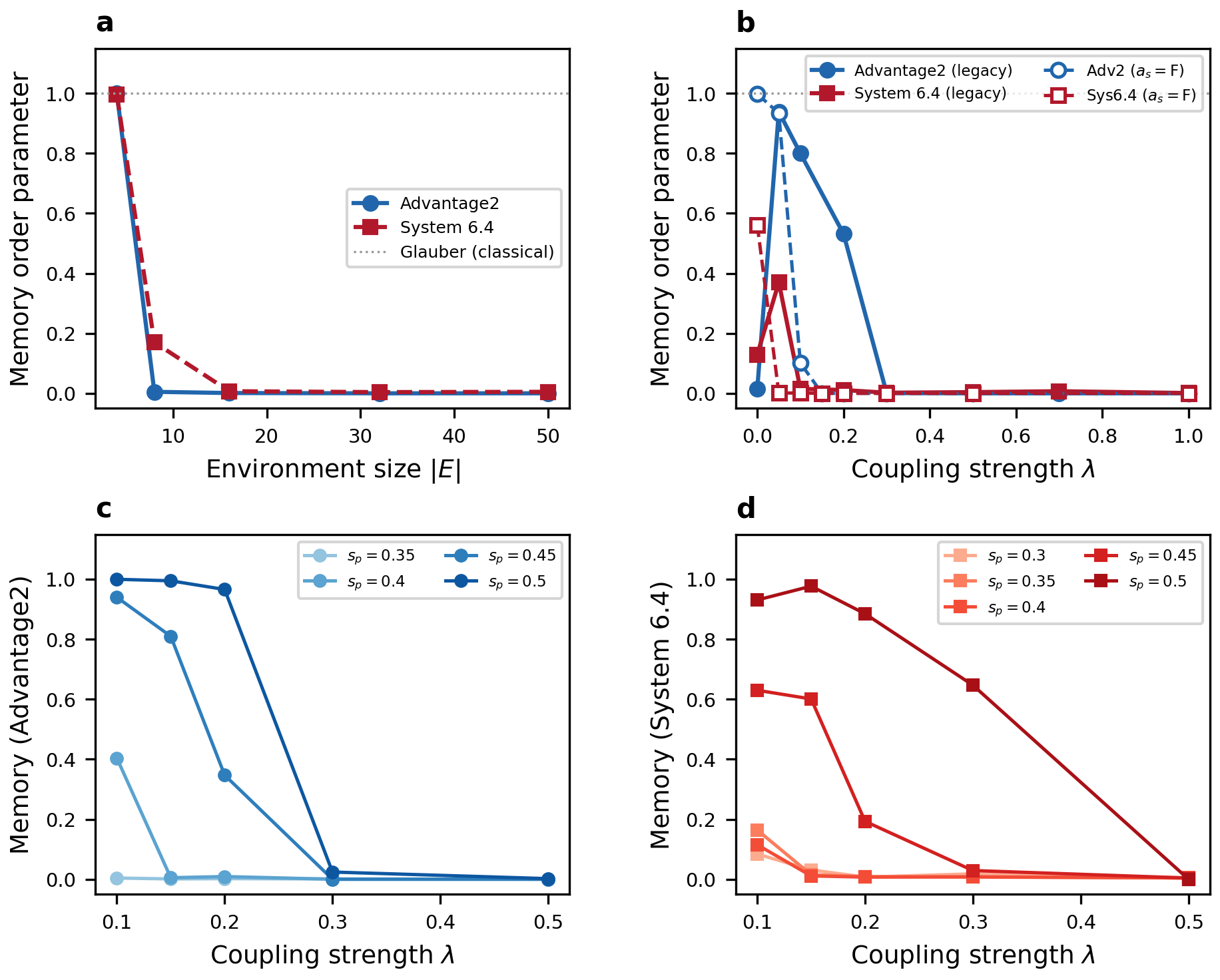}
\caption{\textbf{Bath size and coupling tune a memory-to-relaxation crossover, reproduced on both QPU generations.}
\textbf{a,}~Memory order parameter $\mathcal{M}$ versus environment size $|E|$ on both QPUs ($s_p=0.4$, $\lambda=0.5$).
\textbf{b,}~Memory versus coupling strength $\lambda$ ($|E|=50$).  Filled markers with solid lines: legacy SDK-default submission path.  Open markers with dashed lines: regenerated on direct native-qubit submissions with \texttt{auto\_scale=False} (no chain couplers).  The $\lambda=0$ point is a decoupled control and is excluded from the monotonic crossover statement.
\textbf{c,d,}~Pause-depth sweeps on Advantage2 (\textbf{c}) and System~6.4 (\textbf{d}); lower $s_p$ shifts relaxation to weaker coupling.
Plotted threshold values are nominal SDK-default quantities and inherit the \texttt{auto\_scale} systematic characterized in Methods and the Supplemental Material; bootstrap error bars are smaller than the markers unless shown.}
\label{fig:subsystem}
\end{figure*}

With the protocol and subsystem partition defined, we first show that enlarging and more strongly coupling the environment drives the subsystem toward initial-state independence within experimental resolution on both QPU generations (Fig.~\ref{fig:subsystem}).
On the Advantage2 QPU ($s_p=0.4$, $t_p=100\,\mu$s), the memory order parameter drops sharply with environment size: from $\mathcal{M}=1.0$ at $|E|=4$ to $\mathcal{M}=0.005$ at $|E|=8$ and $\mathcal{M}<0.005$ for $|E|\geq 16$ (Fig.~\ref{fig:subsystem}a, blue circles).
The Advantage\_system6.4 QPU shows the same qualitative trend, with a slightly higher residual memory at intermediate~$|E|$ consistent with its lower energy-to-temperature ratio (Fig.~\ref{fig:subsystem}a, red squares).

The coupling strength~$\lambda$ provides an independent control parameter (Fig.~\ref{fig:subsystem}b).
On Advantage2, memory decreases monotonically from $\mathcal{M}=0.94$ at $\lambda=0.05$ through $0.53$ at $\lambda=0.2$ to $\mathcal{M}<0.001$ for $\lambda\geq 0.3$.
On System~6.4, a similar crossover occurs at lower nominal~$\lambda$, illustrating that the threshold is platform- and submission-path-dependent rather than set by $\beta_\mathrm{eff}\lambda$ alone.
The submission-path dependence is itself quantifiable and reproducible: paired \texttt{auto\_scale} on/off repeats of the same $26$-qubit embedded problem in the transition region shift $\mathcal{M}$ from $\approx 0.99$ to $\approx 0.005$ ($\Delta\mathcal{M}\approx 0.98$, stable across four back-to-back repeats), while saturated conditions away from the crossover are unaffected (Methods; Supplemental Material~\cite{lozano2026supplement})---so all absolute thresholds in this work are reported as empirical quantities for the stated submission path.

To probe the role of the pause-point Hamiltonian, we sweep~$s_p$ on both QPUs (Fig.~\ref{fig:subsystem}c,d).
The crossover curves display the same sigmoidal shape on both platforms: as $s_p$ increases from 0.35 to 0.50, the crossover coupling~$\lambda_c$ shifts to larger values, as both Hamiltonian coefficients and open-system rates change with pause depth.
In~situ single-qubit probes at $s_p=0.4$ yield $\beta_\mathrm{eff}=7.22\pm 0.06$ (Advantage2) and $\beta_\mathrm{eff}=4.29\pm 0.29$ (System~6.4) (Fig.~\ref{fig:disorder}c; see Methods), a ratio of $1.68$; the quoted spreads are population standard deviations across probe estimates, not measurement uncertainties or confidence intervals.
The measured crossover shifts are not captured by the probe slope alone: rescaling the boundary coupling by $\beta_\mathrm{eff}^{\mathrm{S6.4}}/\beta_\mathrm{eff}^{\mathrm{A2}}\approx0.59$ leaves a residual threshold ratio of approximately three at $s_p=0.4$, where $A(s_p)/B(s_p)$ is nearly equal on the two processors ($0.260$ and $0.259$).  The comparison therefore does not yield a single-parameter collapse.  Differences in embedding, calibration and open-system dynamics remain confounded.
The cross-QPU comparison is anchored to a single $(s_p,t_p)=(0.4,\,100\,\mu\mathrm{s})$ operating point on each device, with $t_p$ set well above the relaxation timescale characterized in the kinetics scan; we did not separately sweep $t_p$ at $s_p=0.4$ on Advantage\_system6.4 (Supplemental Material).

\subsection{Environment-state dependence and kinetics}

We next ask which environment states produce this initial-state independence and on what timescale it develops.
At the relaxed working point ($|E|=50$, $\lambda=0.5$), an ordered $E$ (all-up or all-down) gives $\mathcal{M}=0.000$, while a random or domain-wall $E$ leaves residual memory $\mathcal{M}=0.39$ and $0.65$ respectively, with $P_S$ depending on which $E$ preparation was used (cross-preparation $\mathrm{TVD}=1.0$).
The on-chip environment is therefore a finite-size partner whose initial microstate influences the subsystem's fate, not a universal thermal bath.
Pause-time kinetics show that the ordered/disordered distinction persists over the tested interval: with ordered $E$, $\mathcal{M}$ drops from $0.017$ at $t_p=1\,\mu$s to $<0.001$ by $t_p=5\,\mu$s and remains zero to $t_p=500\,\mu$s, while with random $E$ it fluctuates between $0.1$ and $0.7$ over the same range with no systematic decay.
We adopt $t_p=100\,\mu$s ($\sim 20\times$ the ordered-$E$ relaxation timescale) as the working pause time for all subsequent campaigns.
A complementary test using $E$ states produced by forward-annealing the $E$-only Hamiltonian (with $S$ decoupled), which yields low-energy ordered samples, gives $\mathcal{M}=0.000$ across the tested preparations, consistent with sensitivity to the environment's initial energy sector; the preparation also changes its magnetization and boundary fields, so this control does not isolate a thermal-bath mechanism.

\subsection{Disorder as a control parameter for relaxation}

\begin{figure*}[!htbp]
\centering
\includegraphics[width=\textwidth,height=0.55\textheight,keepaspectratio]{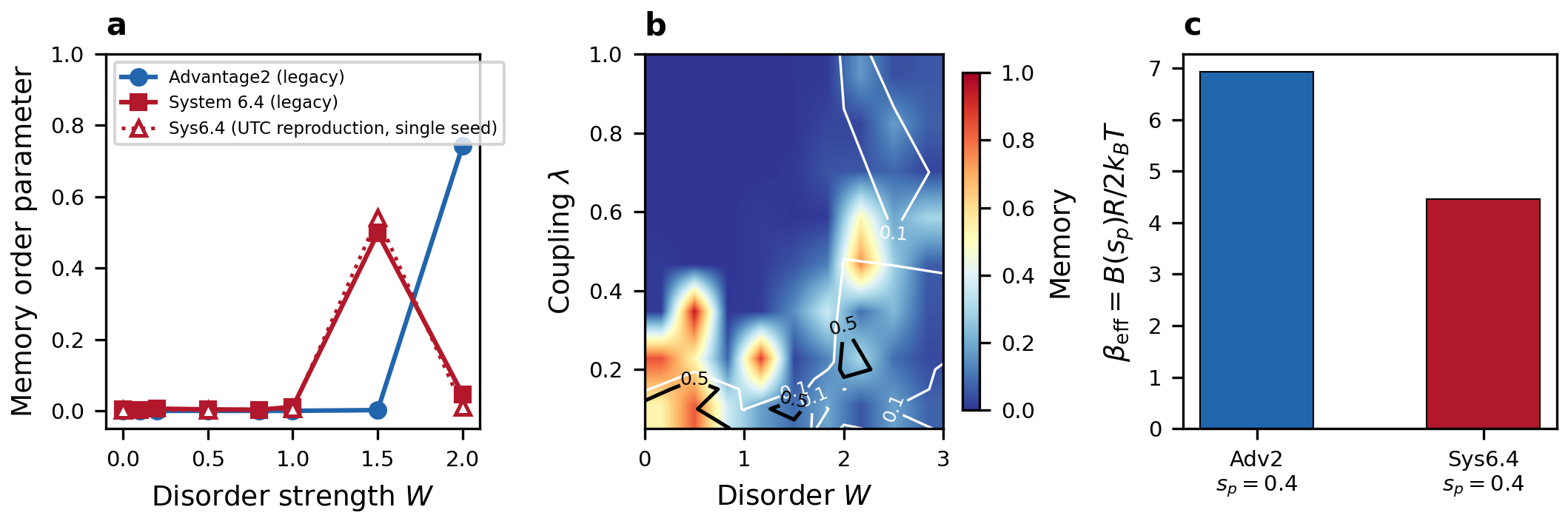}
\caption{\textbf{Disorder degrades relaxation above $W\approx 1.5$.}
\textbf{a,}~Memory order parameter versus disorder strength~$W$ on both QPUs ($|E|=50$, $\lambda=0.5$, $s_p=0.4$).  Solid curves: legacy SDK-default submission path.  Open triangles (dotted, red): a single-seed reproduction on \texttt{Advantage\_system6.4} at the \texttt{uniform\_torque\_compensation} chain-strength path with \texttt{auto\_scale=True}, giving $\mathcal{M}=0.54$ at $W=1.5$ versus the legacy $0.50$.  Memory rises above the noise floor for a subset of realizations beyond $W\approx 1.5$--$2.0$ (5-seed average $\mathcal{M}=0.125\pm 0.103$ at $W=2.0$); this comparison does not isolate the device parameter responsible for the threshold shift.  Smaller fixed chain strengths (Supplemental Material) give weaker arrest, so the arrest magnitude here is specific to the legacy chain-strength regime.
\textbf{b,}~Single-size empirical crossover in the $(\text{disorder}\;W,\;\text{coupling}\;\lambda)$ plane on Advantage2 (72 grid points, $|E|=50$, $|S|=6$).  Color indicates $\mathcal{M}$; contour lines at $\mathcal{M}=0.1$ (white) and $0.5$ (black) delineate the relaxed and memory-retaining regions at this single system size (not a thermodynamic phase diagram).  Stronger coupling pushes the disorder threshold higher.
\textbf{c,}~In~situ measured protocol-level readout slope $\beta_\mathrm{eff}$ at $s_p=0.4$ on both QPUs (error bars: population standard deviations across probe estimates).}
\label{fig:disorder}
\end{figure*}

The sensitivity to the environment's initial energy sector suggests that disorder should compete directly with the same relaxation mechanism, making quenched disorder a third control knob---alongside bath size and coupling---that a sampler user must account for; we test it by adding random longitudinal fields (Fig.~\ref{fig:disorder}a).
At fixed $|E|=50$ and $\lambda=0.5$, $\mathcal{M}$ remains near zero for $W\leq 1.0$ on both QPUs and then rises above the noise floor for a subset of disorder realizations beyond $W\approx 1.5$, reaching single-realization values of $\mathcal{M}=0.74$ at $W=2.0$ on Advantage2 and $\mathcal{M}=0.50$ at $W=1.5$ on System~6.4, without isolating which device difference sets the threshold.
The effect is realization-dependent: a five-realization disorder average on Advantage2 gives $\mathcal{M}=0.000\pm 0.000$ for $W\leq 1.0$, $0.001\pm 0.001$ at $W=1.5$, and $0.125\pm 0.103$ at $W=2.0$, where the dispersion as large as the mean (5 seeds) indicates that some realizations arrest while most do not.
The full $(\lambda, W)$ crossover diagram on Advantage2 (Fig.~\ref{fig:disorder}b, 72 grid points) reveals, at the single system size studied ($|S|=6$, $|E|=50$), an empirical boundary separating initial-state-independent and memory-retaining regions, with stronger coupling pushing the disorder threshold higher; absent a finite-size scaling analysis we treat this as a single-size threshold, not a thermodynamic phase boundary.
The disorder-arrest threshold is graph-dependent: native Zephyr subgraphs, whose boundary connectivity is roughly an order of magnitude denser than the random 3-regular logical graphs studied here, relax across a broader range of $W$ (Supplemental Material).
The lower-connectivity logical graph therefore provides a more stringent test of disorder arrest. We note only in passing that the open-system nature of the QPU and the finite pause time preclude any identification with a many-body-localization transition, despite a superficial resemblance; the operational point is simply that disorder above $W\approx 1.5$ degrades relaxation and must be logged as a sampler-validation parameter.

\subsection{Calibrated conditional-Boltzmann diagnostic}

We now compare the preparation-specific subsystem marginals with a fixed classical conditional-Boltzmann reference calibrated on separate single-qubit probes.  Two questions are kept distinct: whether pooling hides preparation dependence, and whether the chosen environment condition describes the distribution being tested.  The same saved measurements permit a direct comparison between paired memory-and-distance screening and the simpler requirement that each preparation match the reference.

\paragraph{In-situ $\beta_\mathrm{eff}$ calibration.}
Following classical thermometry for quantum annealers~\cite{grattan2025thermometry}, we define $\beta_\mathrm{eff}$ as the Boltzmann slope of the $z$-basis readout distribution produced by the complete reverse-anneal protocol on a single-qubit probe (longitudinal bias $h=0.5$, same schedule $s_p=0.4$, $t_p=100\,\mu$s, \texttt{auto\_scale=False}, 5{,}000 reads per initial state),
\begin{equation}
    \beta_\mathrm{eff}
    = \frac{1}{2h}
      \ln\!\left(\frac{n_\downarrow}{n_\uparrow}\right),
    \label{eq:beta_eff}
\end{equation}
averaged over three probe qubits.
This gives $\beta_\mathrm{eff}=7.219\pm 0.063$ for the main Advantage2 thermal-marginal sweep, $\beta_\mathrm{eff}=7.331\pm 0.198$ for the separately recalibrated frustrated pilot, and $\beta_\mathrm{eff}=4.289\pm 0.294$ for the Advantage\_system6.4 cross-QPU replication.
The quoted spreads are population standard deviations across probe estimates, not standard errors or confidence intervals.
In every case $\beta_\mathrm{eff}$ is a protocol-level readout slope, not a global thermodynamic temperature, and we do not use it to claim a measured temperature of the many-body relaxed state.
As quantified below, sweeping the reference temperature over $\beta\in[\beta_\mathrm{eff}/3,\,3\beta_\mathrm{eff}]$ leaves the relaxed-subset conditional-Boltzmann distance essentially unchanged and well below the conservative finite-sample reference scale, as expected when both distributions are nearly deterministic.
$\beta_\mathrm{eff}$ therefore defines a fixed comparison distribution; it is not a fitted many-body temperature.

\paragraph{Conditional Gibbs prediction.}
For an environment prepared in a fixed ordered state $\sigma_E=\mathbf{1}$, the classical target for the subsystem is the conditional marginal
\begin{equation}
    P_S^\mathrm{th}(\sigma_S \mid \sigma_E = \mathbf{1})
    =
    \frac{\exp[-\beta_\mathrm{eff} H_P(\sigma_S,\sigma_E=\mathbf{1})]}
         {\sum_{\sigma_S'}\exp[-\beta_\mathrm{eff} H_P(\sigma_S',\sigma_E=\mathbf{1})]},
    \label{eq:cond_gibbs}
\end{equation}
evaluated by direct enumeration of the $2^{|S|}$ subsystem configurations with $\sigma_E$ held fixed in the reference.  The physical environment is initialized in this state but is not clamped during evolution; the comparison therefore tests this chosen reference, not an exact prediction for an evolving bath.
This conditional object differs from the unconditional equilibrium marginal, which would average over both ordered $Z_2$ sectors and predict zero subsystem magnetization in globally spin-flip-symmetric instances.

\paragraph{Pooling can conceal preparation dependence.}
Let $P_a=P_S^{(a)}$ denote the measured marginal for preparation $a\in\{\uparrow,\downarrow,\mathrm{random}\}$ and let $Q$ be the reference in Eq.~\eqref{eq:cond_gibbs}.  We compare the equal pooled marginal $\overline{P}=\tfrac13\sum_aP_a$ with the individual marginals through
\begin{equation}
\begin{split}
D_\mathrm{pool}&=\mathrm{TVD}(\overline{P},Q),\\
D_\mathrm{max}&=\max_a\mathrm{TVD}(P_a,Q).
\end{split}
\label{eq:screening_distances}
\end{equation}
Here $D_\mathrm{pool}$ is the quantity denoted $D_\mathrm{TV}$ in the figures and earlier comparisons.  Pooling can cancel discrepancies in the individual marginals.  For a recorded $N=10$, $|S|=4$, $W=1$, $\lambda=0.5$, seed-42 condition on a random 3-regular graph, the probabilities of all four subsystem spins being up are $0.9985$, $0.0005$ and $0.1325$ for the up, down and random preparations.  Their equal mixture gives $0.377167$, close to the reference probability $0.368566$.  The full pooled distance is $D_\mathrm{pool}=0.008601$, even though $\mathcal{M}=0.998$ and $D_\mathrm{max}=0.629934$.  All three preparation-specific marginals fail the $0.05$ distance cutoff; their distances are $0.629934$, $0.370173$ and $0.238173$.  The Supplemental Material shows the corresponding four-bar worked example and provides the exact saved marginals.

\paragraph{Comparison of screening rules.}
At a common cutoff $t$, the pooled rule selects $D_\mathrm{pool}<t$; the paired rule additionally requires $\mathcal{M}\le t$; the per-preparation rule selects $D_\mathrm{max}<t$.  Recomputing all three rules over both campaigns gives Table~\ref{tab:screening}.  At $t=0.05$, the paired and per-preparation rules select exactly the same $110$ Advantage2 conditions and the same $22$ System~6.4 conditions, verified by condition identity as well as count.  Compared with pooling alone, either rule excludes $22$ conditions on Advantage2 and $10$ on System~6.4.  These counts describe the archived acquisition conditions; they do not measure accuracy against independently established equilibrium labels.

\begin{table}[!htbp]
\centering
\caption{\label{tab:screening}Conditions selected by three screening rules at $t=0.05$.  The paired and per-preparation selections have identical membership on each processor.  Each rule uses the same saved preparation marginals and the same calibrated, initial-boundary reference.}
\begin{ruledtabular}
\begin{tabular}{lrr}
Rule & Advantage2 & System 6.4\\
\colrule
Pooled distance & $132/494$ & $32/60$\\
Memory and pooled distance & $110/494$ & $22/60$\\
Each preparation separately & $110/494$ & $22/60$\\
\end{tabular}
\end{ruledtabular}
\end{table}

The equality of selections is empirical, not a general equivalence of the rules.  For a common fixed reference $Q$, convexity and the triangle inequality give
\begin{equation}
D_\mathrm{pool}\le D_\mathrm{max},\qquad
\mathcal{M}\le2D_\mathrm{max}.
\label{eq:screening_bounds}
\end{equation}
Thus $D_\mathrm{max}<t/2$ guarantees the paired conditions at $t$, but $D_\mathrm{max}<t$ does not generally imply $\mathcal{M}\le t$.  At $t=0.02$ the observed memberships again agree on both processors; at $t=0.10$ they agree on Advantage2 and differ by one condition on System~6.4 (Supplemental Material, Preparation-resolved screening and pooling cancellation).  The observable pair is useful for distinguishing initialization dependence from disagreement with a specified target, without being mathematically necessary or sufficient for general sampler validation.  In particular, $\mathcal{M}$ remains measurable when no trusted reference is available, whereas either distance rule requires one.

\begin{figure*}[!htbp]
\centering
\includegraphics[width=\textwidth,height=0.55\textheight,keepaspectratio]{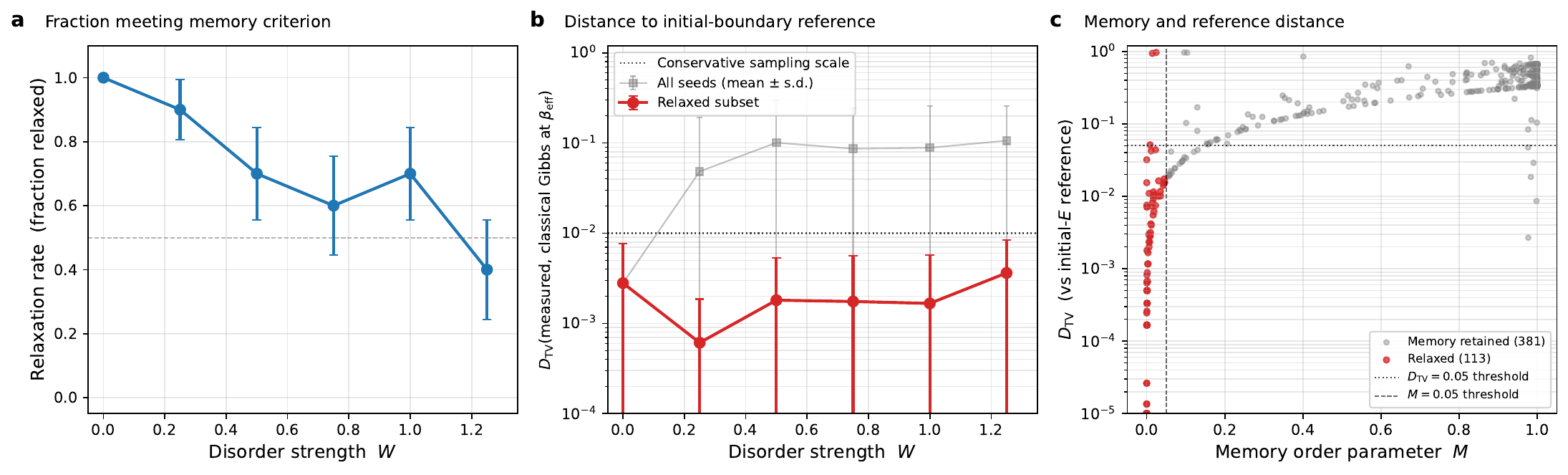}
\caption{\textbf{Near-deterministic relaxed readouts agree with the chosen conditional reference, while two globally reversed readouts pass the memory test yet miss the initial-boundary reference by $D_\mathrm{TV}>0.94$.}
\textbf{a,}~Fraction of relaxed disorder realizations ($\mathcal{M}\le 0.05$) in the 10-seed disorder sweep ($N=12$, $|S|=4$, $\lambda=0.5$, 3-regular logical graph, Advantage2).  Relaxation fraction decreases from $10/10$ at $W=0$ to $4/10$ at $W=1.25$.
\textbf{b,}~Total-variation distance between the pooled subsystem marginal and the calibrated classical conditional Gibbs target at $\beta_\mathrm{eff}=7.219$ vs disorder strength.  Open circles: mean over all 10 seeds; filled circles: mean over the relaxed subset only (error bars = s.d. across relaxed seeds).  Gray band: conservative independent-sampling reference scale $D_\mathrm{TV}\simeq\tfrac12\sqrt{K/N_\mathrm{reads}}\approx 0.026$ with $N_\mathrm{reads}=6000$, $K=2^{|S|}=16$; the relaxed subset stays at or below this scale across the sweep.
\textbf{c,}~Scatter of $D_\mathrm{TV}$ vs $\mathcal{M}$ across all thermal-marginal conditions.  Red: relaxed ($\mathcal{M}\le 0.05$); gray: memory-retaining.  Two top-left red outliers correspond to the same $\lambda=0.2$, $N=8$, $W=1$, seed~42 native Zephyr instance at Hamiltonian scales $0.35$ and $0.5$ (final-environment reversal, discussed in the text).  Distances below $10^{-5}$ are shown at the lower display limit.  The two criteria give the four-cell counts $110/3/22/359$ (text); the three relaxed distance failures comprise the large-discrepancy pair and one borderline $D_\mathrm{TV}=0.051$ miss.  Retrospective checks and the plotting-script revision used AI assistance (Methods).}
\label{fig:thermal_marginal}
\end{figure*}

\paragraph{Relaxed-subset distances are a consistency check, not a thermometric test.}
In a 10-seed disorder sweep at $N=12$, $|S|=4$, $\lambda=0.5$ on a 3-regular logical graph (Fig.~\ref{fig:thermal_marginal}), relaxed conditions sit very close to the conditional classical reference: mean $D_\mathrm{TV}$ over relaxed realizations lies between $0.6\times 10^{-3}$ and $3.6\times 10^{-3}$ across $W\in\{0,0.25,0.5,0.75,1.0,1.25\}$, with no monotone $W$ dependence (Fig.~\ref{fig:thermal_marginal}b), while the relaxed fraction decreases from $10/10$ at $W=0$ to $4/10$ at $W=1.25$ (Fig.~\ref{fig:thermal_marginal}a).
These distances are below the conservative distribution-independent scale $\tfrac12\sqrt{K/n}\approx0.026$ for $K=16$ states and $n=6000$ reads.  This scale is an upper-bound estimate of expected empirical total-variation error under independent sampling, not a lower noise floor; near-deterministic distributions can have much smaller sampling error.  The measured marginals and references are concentrated on the prepared-boundary configuration; we therefore read this as a consistency check, not as a stringent confirmation of the Boltzmann \emph{form}.
A direct test makes this explicit: sweeping the reference inverse temperature over $\beta\in[\beta_\mathrm{eff}/3,\,3\beta_\mathrm{eff}]$ changes the relaxed-subset median $D_\mathrm{TV}$ only between $1.7\times 10^{-4}$ and $3.5\times 10^{-3}$---remaining below that conservative reference scale---so the agreement does not localize a temperature (Supplemental Material, $\beta$-sensitivity).
Across the full $494$-condition Advantage2 sweep (Zephyr-derived graphs, random 3-regular, three Hamiltonian scales, large-bath native and disorder sub-experiments), $113/494$ conditions are relaxed; all but three have conditional $D_\mathrm{TV}<0.05$, and the three exceptions are two large initial-boundary discrepancies ($D_\mathrm{TV}>0.94$) and one borderline miss described below, while the $381$ memory-retaining conditions sit far from the reference (median $D_\mathrm{TV}=0.35$).
A $60$-condition cross-QPU replication on Advantage\_system6.4 at the separately calibrated $\beta_\mathrm{eff}=4.29$ gives $22/60$ relaxed conditions with conditional $D_\mathrm{TV}\in[10^{-7},\,0.029]$ (median $0.005$).
The campaign spans graph geometry, subsystem size $|S|\in\{4,6\}$, coupling $\lambda\in\{0.2,0.3,0.5\}$ and Hamiltonian scale (Fig.~\ref{fig:thermal_marginal}c).  Its small relaxed-subset distances do not establish a thermometrically calibrated Boltzmann state.

\paragraph{Quantum reduced Gibbs diagnostic.}
The same comparison can be repeated with the $z$-basis diagonal of the
quantum reduced Gibbs state of the pause-point Hamiltonian,
$H(s_p) = -\tfrac{A(s_p)}{2}\sum_i \sigma_x^i + \tfrac{B(s_p)}{2} H_P$,
evaluated on the full $N$-qubit system by exact diagonalization for
$N\le12$, with the transverse field retained at $s_p$.  We use the same
protocol-level $\beta_\mathrm{eff}$ and the device-calibrated ratio
$A(s_p)/B(s_p)=0.260$ for Advantage2 and $0.259$ for
Advantage\_system6.4, extracted from the solver anneal-schedule
spreadsheets at $s_p=0.4$.  Two quantum targets are compared against the
measured marginal: the \emph{conditional} quantum reduced Gibbs marginal
with $\sigma_E$ fixed to the prepared ordered state, and the
\emph{unconditional} marginal tracing over all $2^{|E|}$ environment
configurations.
The unconditional target gives median $D_\mathrm{TV}\approx0.5$ in the Advantage2 disorder sweep.  At $W=0$ it averages over both spin-flip-related ordered sectors; random longitudinal fields at $W>0$ break that symmetry, but the unconditional target still does not condition on the prepared all-up environment; the conditional target
agrees on the disorder sweep (median $D_\mathrm{TV}=0.0145$ across 43
relaxed Advantage2 conditions, $0.0146$ across 22 Advantage\_system6.4
conditions) but the classical conditional target is uniformly tighter.
The reference is constructed at the pause point, whereas the measurement follows the return ramp; a tighter classical fit to nearly deterministic readouts does not identify when or how that difference arose.  Full per-condition $D_\mathrm{TV}$ tables and the
Hamiltonian-scale dependence of the conditional quantum target are
reported in the Supplemental Material.

\paragraph{Large reference discrepancies and evolution of the environment.}
Three relaxed conditions have $D_\mathrm{TV}>0.05$ against the initial-boundary classical target.  The two largest ($0.9409$ and $0.9674$) are one $N=8$, $\lambda=0.2$, $W=1$, seed-42 instance at scales $0.35$ and $0.5$; the third is a borderline seed-123 condition (Supplemental Material).
A retrospective offline check reconstructed each of the $6000$ full-system configurations per scale by matching its stored energy to exact enumeration of the recorded Hamiltonian.  Every read matched a unique configuration at tolerance $10^{-7}$, and each sampleset admitted a consistent column permutation reproducing the raw spins; the reconstructed subsystem marginals exactly matched the stored summaries.
At both scales the observed final environment was never all-up.  The dominant all-down configuration, with weights $0.9918$ and $0.9853$, is the unique global ground state, lower than the all-up configuration by $1.9401$ and $2.7716$ in the recorded coupling units.  Thus the large distance to the all-up-boundary reference does not demonstrate trapping in a higher-energy basin.
As an exploratory sensitivity check, replacing the initial boundary by a mixture of conditional classical targets weighted by the observed final environment reduces $D_\mathrm{TV}$ to $0.1253$ and $0.0598$.  The corresponding full-system classical-Gibbs marginal gives $0.1239$ and $0.0572$ at the same probe slope.  Both revised distances remain above $0.05$.  The final-environment mixture uses the measured data and is not an independent validation target; neither comparison establishes equilibration or a many-body temperature.
The two observables should therefore be interpreted as memory and agreement with a specified reference.  Assigning a physical failure mode requires checking that reference's conditioning assumptions.

At the original cutoffs, the full Advantage2 campaign has $110$ conditions passing both memory and pooled-distance criteria, $3$ passing memory alone, $22$ passing pooled distance alone, and $359$ passing neither.  The two large initial-boundary discrepancies persist at $D_\mathrm{TV}\ge0.85$ when the reference slope is varied over $[\beta_\mathrm{eff}/3,3\beta_\mathrm{eff}]$.  Changing the reference temperature therefore does not repair the boundary mismatch.  Their distance classification also persists over the tested $0.02$--$0.10$ cutoffs, although their memory classification can change (Supplemental Material).  Checking the reference's conditioning assumptions is necessary before assigning a physical failure mode to a large distance.

\subsection{Classical and small-system controls}

\begin{figure*}[!htbp]
\centering
\includegraphics[width=\textwidth,height=0.55\textheight,keepaspectratio]{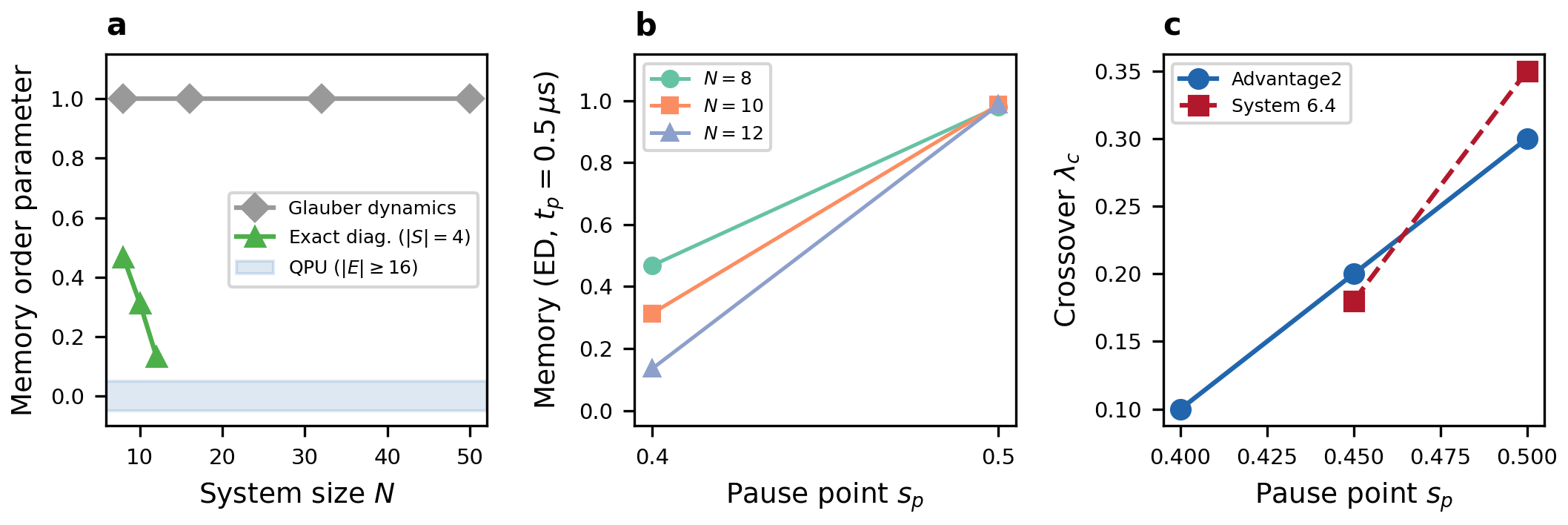}
\caption{\textbf{Finite-time Glauber and small-system ED controls.}
\textbf{a,}~Archived Glauber and closed-system ED memory at $s_p=0.4$, $t_p=0.5\,\mu$s; the blue band marks the QPU observation for $|E|\ge16$.  Preparation, system size and evolution duration differ between these controls and the QPU, and the Glauber configuration predates the probe calibration (text).
\textbf{b,}~Closed-system ED memory versus pause depth for three system sizes at $t_p=0.5\,\mu$s.
\textbf{c,}~Approximate nominal crossover couplings read from the legacy QPU pause-depth sweep.  These values inherit the SDK-default auto-scale systematic and the joint variation of $A$, $B$ and open-system rates with $s_p$.  The plotting-script revision used AI assistance (Methods).}
\label{fig:baselines}
\end{figure*}

Figure~\ref{fig:baselines} places the hardware observations alongside finite-time Glauber and small-system closed quantum calculations.
The archived Glauber calculations give $\mathcal{M}=1$ for $N\in\{8,16,32,50\}$, two random seeds and two temperature labels; the corresponding driver initializes the entire $S{+}E$ system in all-up, all-down or random states, unlike the QPU protocol, which changes the preparation of $S$ while retaining the same initial preparation of $E$.
The configuration specifies dimensionless temperatures $0.112$ and $0.221$, whereas the later probe calibrations give $T_\mathrm{cal}=0.139$ and $0.233$.
Because preparation and temperature are unmatched, these calculations characterize the chosen updates and do not isolate a quantum relaxation mechanism or establish equilibrium mixing on the QPU.
The archived closed-system ED calculation for $N\in\{8,10,12\}$ ($|S|=4$), $s_p=0.4$ and $t_p=0.5\,\mu$s gives $\mathcal{M}=0.468$, $0.313$ and $0.134$, respectively; at $s_p=0.5$ the values exceed $0.98$.
This is a comparison within a closed-system model, not a time- or size-matched simulation of the hardware experiment.
The repository also contains a phenomenological Lindblad solver and an isotropic $O(3)$ model, which differs from the transverse-field Ising spin-vector Monte Carlo model of Ref.~\cite{shin2014svmc}.  Their campaign outputs are not archived, so they do not enter the empirical comparison (Supplemental Material).
On the QPU, changing $s_p$ changes $A(s_p)$, $B(s_p)$ and open-system rates simultaneously, so the observed pause-depth dependence does not separately identify their contributions.

\section{Discussion}

Preserving preparation labels changes which reverse-anneal settings pass a reference test.  Of the $132$ Advantage2 conditions with pooled distance below $0.05$, $22$ retain memory above $0.05$; on the second processor the corresponding count is $10$ of $32$.  The worked example shows how this happens: preparations concentrated in different regions can average to a marginal close to the chosen target.  Requiring each preparation separately to pass removes the same conditions as the paired memory-and-pooled-distance rule at the stated cutoffs (Table~\ref{tab:screening}).  These are screening results within the archived campaigns, not estimates of sampler accuracy against an independent equilibrium classification.

Memory nevertheless carries information beyond that decision.  It can be measured when the target is unknown, computationally inaccessible or uncertain because its boundary assumption is suspect.  A reference distance then addresses agreement with a specified distribution, while the preparation-resolved distances reveal whether pooling hides a disagreement.  The observables support interpretation; their agreement does not certify an arbitrary sampler.

The boundary reconstruction gives a second operational lesson.  Of the $113$ Advantage2 conditions with $\mathcal{M}\le0.05$, three miss the initial-boundary reference: two large discrepancies use one seed-42 instance at scales $0.35$ and $0.5$, while a borderline miss ($D_\mathrm{TV}=0.051$) uses seed~123 at scale $0.35$.  The seed-42 outputs concentrate in the global ground state after the environment reverses.  Their original distances reproduce, but interpreting those distances as higher-energy trapping would mistake the initial boundary for a constraint on the physical evolution.  The final-environment comparisons remain above the $0.05$ cutoff and establish neither equilibration nor a many-body temperature.

D-Wave QPUs operate as open quantum systems with cryogenic bath, flux noise and time-dependent anneal schedule all participating in the dynamics~\cite{albash2018adiabatic,mehta2025dwave,mehta2025unraveling,aronoff2026coherence}.
The relaxation threshold in $|E|$ characterises the effective boundary-coupling density of the chosen graph family rather than a universal qubit-count threshold: native Zephyr subgraphs, with roughly an order of magnitude denser boundary connectivity, relax at smaller $|E|$ on the same logical problem than the random 3-regular graphs in Fig.~\ref{fig:subsystem}a; conversely at $|E|=20$, $\lambda=0.5$ a direct native-qubit submission remains memory-retaining while the embedded random-3-regular submission at larger $|E|=50$ relaxes (Supplemental Material).
The small relaxed-subset distances also occur in the Zephyr-graph analysis (mean conditional $D_\mathrm{TV}=0.0015$ across $5$ relaxed conditions, $N=12$, $|S|=4$, $\lambda=0.5$), although this graph-family comparison alone does not isolate the physical embedding.

For annealer-based Boltzmann-machine training~\cite{adachi2015qbm,amin2018qbm,benedetti2017qbm}, quantum-assisted sampling and reverse-anneal optimization~\cite{marshall2019pausing,chen2020pausing,pelofske2025erasing,nelson2022gibbs,vuffray2022boltzmann,grattan2025thermometry}, the measurements support retaining preparation-specific distributions and reporting their memory before pooling.  When a target is available, each preparation can be tested against it; when it is conditional on an environment state, that condition must be justified.  Submission metadata (Table~\ref{tab:metadata}) belongs with the samples because changing the programming path can change the classification.  These recommendations concern validation of the measured distributions; no downstream learning or optimization improvement was measured here.
\begin{table*}[!htbp]
\centering
\caption{\label{tab:metadata}Submission-path metadata and its measured impact on the validation observables in this work (spot checks, not a factorial design; details in Methods and the Supplemental Material).}
\begin{ruledtabular}
\begin{tabular}{ll}
Metadata knob & Measured impact \\
\colrule
\texttt{auto\_scale} flag & $\Delta\mathcal{M}$ up to $\approx 0.98$ (transition region); $\le 0.11$ saturated \\
Chain-strength path & disorder-arrest $\mathcal{M}=0.54$ (legacy UTC) vs $\lesssim 0.1$ (fixed $c\in\{1,2\}$) \\
Native vs embedded & $\lambda_c\in[0.05,0.10]$ native vs $[0.15,0.20]$ embedded ($c=1$) \\
Environment preparation & $\mathcal{M}=0.000$ ordered vs $0.39$ random, $0.65$ domain wall \\
Spin-reversal gauge & $\sigma_\mathcal{M}\le 0.02$ across four gauges \\
Return-ramp duration & classification unchanged, $t_r\in\{1,5,20\}\,\mu$s \\
\texttt{reinitialize\_state} & \texttt{True} read-count-independent; \texttt{False} drifts with reads \\
Reference $\beta$ ($9\times$ range) & large initial-boundary discrepancies persist \\
\end{tabular}
\end{ruledtabular}
\end{table*}
Environment preparation is part of the sampling protocol.  In the tested ferromagnetic graph family, ordered preparations lead to memory loss, whereas random and domain-wall preparations leave residual memory.  The preparation should therefore be specified and its final-state evolution checked when choosing a conditional reference; these observations do not prescribe an environment state for an arbitrary target distribution.
The question of whether an annealer can act as its own thermal bath~\cite{kendon2026qacm} motivates subsystem experiments, but the present observations do not answer it.  The on-chip environment influences relaxation while an external cryogenic bath and the full time-dependent schedule remain active; initial-state independence and agreement with a prepared-boundary reference do not isolate internal thermalization or establish a universal bath.

The subsystem partition provides a programmable setting for studying relaxation at sizes beyond exact diagonalization, alongside other experimental approaches using cold atoms and trapped ions~\cite{kaufman2016quantum,smith2016mbl,shaw2025ergodicity}.  The present evidence supports a narrower methodological conclusion: a reference test should preserve preparation information, and its physical interpretation depends on the reference's environment assumption.  The memory diagnostic remains available without a trusted target.  Its use transfers across the two processors studied here, whereas the two large initial-boundary discrepancies were observed only on Advantage2.
\section{Methods}

\subsection{Quantum processing units}

Experiments were performed on two D-Wave QPUs~\cite{dwave2024system_docs}: the Zephyr-topology Advantage2 (4{,}579 active qubits, operated as \texttt{Advantage2\_system1.13} for the sweeps of Figs.~\ref{fig:subsystem}--\ref{fig:disorder} and~\ref{fig:baselines}, then \texttt{Advantage2\_system1} with qubit 4374 removed for the thermal-marginal campaign of Fig.~\ref{fig:thermal_marginal}) and the Pegasus-topology Advantage\_system6.4 (5{,}612 active qubits) for the cross-QPU replication.
The dimensionless protocol-level readout slope $\beta_\mathrm{eff}$ was measured in~situ by submitting single-qubit probes at $h=0.5$ under the full reverse-anneal protocol ($s_p=0.4$, $t_p=100\,\mu$s, \texttt{auto\_scale=False}, 5{,}000 reads per initial state), extracting $\beta_\mathrm{eff}=\ln(n_\downarrow/n_\uparrow)/(2h)$ averaged over three probe qubits from the solver nodelist.  We write $T_\mathrm{cal}=1/\beta_\mathrm{eff}$ for the corresponding calibrated readout temperature in the dimensionless coupling units (not a refrigerator temperature); the historical baseline temperatures and their difference from $T_\mathrm{cal}$ are stated separately below.  An identification with $B(s_p)R/(2k_BT)$ would additionally require equilibration at the pause and preservation of that distribution through the return ramp; neither assumption is established here.
This procedure gives $\beta_\mathrm{eff}=7.219\pm 0.063$ for the Advantage2 thermal-marginal campaign and $\beta_\mathrm{eff}=4.289\pm 0.294$ for the Advantage\_system6.4 cross-QPU replication; the frustrated pilot (Supplemental Material; deposited but not used for any main-text conclusion) uses an independently recalibrated $\beta_\mathrm{eff}=7.331\pm 0.198$ from a separate run.  The archived Advantage2 point value $7.219$ sets the reference in every thermal-marginal comparison; all quoted calibration spreads are population standard deviations across probe estimates, not standard errors or confidence intervals.  The original Advantage2 probe-count file is absent from this repository; the $\pm 0.063$ is retained in campaign logs, and an independent $h=0.25$/$h=0.5$ probe comparison spans $7.14$--$7.51$ ($\sim\!5\%$; Supplemental Material).  That bracket measures bias sensitivity rather than a confidence interval for the original campaign.
$\beta_\mathrm{eff}$ is a protocol-level readout slope, not a refrigerator temperature: each value is a single per-campaign single-qubit-probe calibration, not verified for cross-calibration-epoch stability, and it is used only to define a \emph{fixed} conditional-Boltzmann reference for the discrepancy detector, not as a fitted many-body temperature.  The relaxed-subset distance to this reference is essentially independent of $\beta_\mathrm{eff}$ over a ninefold range (Calibrated conditional-Boltzmann diagnostic subsection; Supplementary $\beta$-sensitivity), so the consistency and severe-discrepancy observations do not require a sharply identified reference temperature; this sensitivity check does not measure hardware drift between sessions.
The per-QPU anneal-schedule coefficients used in the quantum reduced Gibbs diagnostic were extracted from the solver schedule spreadsheets~\cite{dwave2024system_docs}, giving $A(s_p)/B(s_p)=0.260$ for Advantage2 and $0.259$ for Advantage\_system6.4 at $s_p=0.4$.
Solver identifiers, qubit availability and the retained per-campaign metadata are described in the Supplemental Material and in the deposited archive.

\subsection{Hamiltonian family and embedding}

The problem Hamiltonian is a random 3-regular Ising model: $H_P = \sum_i h_i \sigma_z^i + \sum_{\langle ij \rangle} J_{ij}\sigma_z^i\sigma_z^j$, with internal couplings $J_{ij}=-1$ (ferromagnetic), boundary couplings between $S$ and $E$ scaled by $\lambda\in[0,1]$, and quenched disorder $h_i\sim\mathcal{U}[-W,W]$.
The main environment-size, coupling, disorder and pause-depth sweeps (Figs.~\ref{fig:subsystem}--\ref{fig:baselines}) use $|S|=6$ with $|E|$ up to $50$; the thermal-marginal campaign uses $|S|=4$ small-system instances at $N\le 12$ to permit exact classical and quantum reference computations.  $S$ is a connected subgraph in all cases.
Logical graphs were embedded with \texttt{EmbeddingComposite}~\cite{cai2014minor} at the default \texttt{uniform\_torque\_compensation} chain strength; chain-break effects, embedding-induced sampling bias and the Zephyr-vs-Pegasus chain-strength differences are discussed in~\cite{marshall2020embedding} and in the Supplemental Material.

\subsection{Reverse-anneal protocol}

All experiments use a four-point reverse-anneal schedule~\cite{dwave2024reverse_anneal}: $(0, 1.0) \to (5\,\mu\mathrm{s}, s_p) \to (5+t_p\,\mu\mathrm{s}, s_p) \to (10+t_p\,\mu\mathrm{s}, 1.0)$, with \texttt{reinitialize\_state=True}.
Default parameters: $s_p=0.4$, $t_p=100\,\mu$s, 1{,}000--2{,}000 reads per configuration.
Unless otherwise stated, initial states of $S$ were varied (all-up, all-down, random) while $E$ was initialized all-up; the environment-preparation controls (all-down, random, domain wall, forward-annealed) and the bath-ensemble sub-experiments use the preparations described in the Results and Supplemental Material.
The submission path differs across sub-experiments and is important for the interpretation of absolute threshold values.
The single-qubit calibration and the explicitly direct native-qubit spot checks use \texttt{auto\_scale=False}.  For the thermal-marginal graph-family campaigns, the archived driver remaps logical labels and defaults to an embedding composite, but the job files omit embedding maps and auto-scale flags.  We therefore distinguish the recorded nominal Hamiltonian and logical topology from unverified per-job physical embedding metadata.
Discovery sweeps and the main subsystem-bath sweeps (Figs.~\ref{fig:subsystem}--\ref{fig:baselines}) were submitted through the SDK-default path with the minor-embedding composite and automatic Hamiltonian rescaling (\texttt{auto\_scale=True}) both active~\cite{dwave2024system_docs}.
We characterized this rescaling through paired \texttt{auto\_scale}-on/off spot-checks at six embedded and unembedded working points, a transition-region $\lambda$-scan, and four independent random spin-reversal gauges (Supplemental Material).
Outside the transition region the two settings agree exactly or within $|\Delta\mathcal{M}|\lesssim 0.11$, with most saturated points agreeing at $0$ or $1$, and gauge-averaged $\mathcal{M}$ has standard deviation $\le 0.02$ across gauges at every tested condition, so the relaxation classification is robust to the submission path away from the crossover.
Inside the transition region the rescaling shifts the effective crossover in $\lambda$ in a reproducible manner, so the absolute $\lambda_c$ values in Figs.~\ref{fig:subsystem} and \ref{fig:baselines} are reported as empirical quantities for the SDK-default submission path used in the legacy campaigns, not as universal numbers; the public code release defaults to \texttt{auto\_scale=False}.

\subsection{Observables and statistical analysis}

The marginal distribution $P_S(\sigma_S) = \sum_{\sigma_E} P(\sigma_S, \sigma_E)$ is computed from QPU samples by projecting onto the subsystem qubit indices, using the variable ordering returned by the sampler.
The memory order parameter is $\mathcal{M} = \max_{a,b}\,\mathrm{TVD}(P_S^{(a)}, P_S^{(b)})$, where $\mathrm{TVD}(P,Q) = \frac{1}{2}\sum_x |P(x)-Q(x)|$.
Read-level bootstrap standard deviations (200 resamples) are retained for the environment-size sweep; these are conditional on the submitted job and its preparation.  They do not account for calibration drift or establish read independence.  Disorder averages use the realization as the independent unit and report the dispersion across the stated seeds.  The deposited phase-4 record contains summary statistics rather than the reads or bootstrap resample arrays, so those uncertainties cannot be reconstructed from raw reads in the present deposit.

\subsection{Retrospective reanalysis}

All reanalyses use retained measurements and require no new QPU execution.  For the screening comparison, a standalone implementation enumerates the conditional classical target from each archived Hamiltonian and probe slope, forms the equal mixture of the three saved preparation-specific marginals, and recalculates $\mathcal{M}$, $D_\mathrm{pool}$ and $D_\mathrm{max}$.  Recomputed memory and pooled distance agree with the archived values to within $10^{-12}$ for all $494+60$ conditions.  The three rules are compared by selected condition identities, as well as counts, at cutoffs $0.02$, $0.05$ and $0.10$; these descriptive thresholds are not fitted to independent equilibrium labels.  The inventories contain acquisition conditions, with instances and seeds reused across parameter settings, rather than $554$ independent instances.

The full-system boundary check is a separate reconstruction from stored read energies, Hamiltonians and spin arrays.  It uses exact energy enumeration to identify configurations and verifies the resulting per-job spin-column permutation and saved subsystem marginals; its recoverable scope is stated in Data availability.  It does not infer equilibration from endpoint energies.

AI tools assisted in writing the retrospective analysis and plotting code. The standalone screening verifier enumerates each reference independently of the original analysis routines and recomputes memory and distance from the saved preparation marginals. This verifies the summary-level calculation. The limits of raw-read reconstruction are given below.

\subsection{Classical baselines}

\emph{Glauber dynamics:} single-spin-flip Monte Carlo with acceptance probability $p = [1+\exp(\Delta E/T)]^{-1}$ in dimensionless coupling units.  The archived configuration specifies $T=0.112$ and $0.221$, 10{,}000 initial sweeps and 2{,}000 subsequent samples, one per sweep.  Autocorrelation and equilibration were not certified for the retained campaign.  Memory retention measures finite-run basin persistence and does not establish a stationary equilibrium distribution.
\emph{Exact diagonalization:} the transverse-field Ising Hamiltonian $H(s_p)$ was constructed as a sparse matrix for $N\leq 12$ and time-evolved via \texttt{expm\_multiply} (SciPy).  The schedule coefficients $A(s_p)$ and $B(s_p)$ in GHz are absorbed into $H(s_p)$ directly, and the physical pause time $t_p$ (in $\mu$s) is converted to the dimensionless time argument of $\exp(-iHt)$ via $t_\mathrm{nat}=2\pi\times 10^{3}\,t_{\mu s}$, where the factor of $2\pi\times 10^3$ reconciles GHz with inverse microseconds (in units where $\hbar=1$).
Subsystem marginals were obtained by tracing out environment degrees of freedom.
\emph{Continuous-spin code:} isotropic O(3) Heisenberg spins with Metropolis updates on the unit sphere, using $H = \sum_i h_i n_i^z + \sum_{ij} J_{ij}(\mathbf{n}_i\cdot\mathbf{n}_j)$ and $z$-axis projected measurement.
The Supplemental Material states the configuration and provenance limits of these calculations.

\subsection{Data availability}

The retained data and original code are openly available at the archived repository commit cited in Ref.~\cite{lozano2026dataset}, also accessible through \url{https://github.com/LuisLozanoM/reverse-anneal-diagnostic}.
The deposit contains raw thermal-marginal D-Wave samplesets (HDF5, with per-read spins, energies, schedule and initial-state attributes), the corresponding Hamiltonians and processed subsystem marginals, calibration summaries and anneal-schedule spreadsheets, and legacy figure summaries.
The memory and classical conditional-distance results for the 494-condition Advantage2 campaign and the 60-condition second-processor campaign can be recomputed from the retained Hamiltonians and per-initial-state marginal summaries without QPU access.  Raw HDF5 files omit the sampler variable ordering.  Retrospective energy enumeration recovers $1380$ of $1620$ tested small-system samplesets and exactly reproduces their saved marginals; degeneracies leave $240$ unresolved, and the $N=56$ conditions were not enumerated (Supplemental Material).
The earlier phase-4 sweeps and classical controls are retained as processed summaries; their raw reads, baseline trajectories and bootstrap resample arrays are not included.  Some HDF5 metadata fields are incomplete: calibration epochs are null, and the recorded \texttt{graph\_id} values are solver identifiers rather than independently verified topology digests.
The original Advantage2 probe-count file is not deposited here; the quoted $\pm0.063$ remains a historical between-probe spread.
The retrospective reanalysis scripts and result files accompany this version as ancillary material.  In particular, the self-contained \texttt{screening} resource includes the Hamiltonians and preparation-specific marginals for all $554$ conditions; it reproduces the screening comparisons without repository access, raw HDF5 files or QPU access.  The separate boundary and temperature-sensitivity analyses use the archived repository data as inputs.

The companion study~\cite{lozano2026cycled} examines schedule-dependent subsystem occupation under cycled reverse annealing.  It shares experimental context and preliminary pilot history with this work; the present article addresses reference validation, and no cross-study pilot comparison supports its conclusions.

\subsection{Code availability}

The repository cited in Data availability contains the reverse-anneal driver, classical comparison codes, subsystem analysis and original figure scripts, with minimum dependencies recorded in \texttt{pyproject.toml}. An exact historical environment lockfile is unavailable. The ancillary code includes the preparation-resolved verifier and plotting script, reference-slope sensitivity analysis, final-environment reconstruction and their outputs. The code is available under Apache-2.0 and the data under CC-BY-4.0.

\subsection*{Version note}

This version revises the interpretation in arXiv:2605.19381v2.  Full-system reconstruction shows that the two large prepared-boundary discrepancies accompany environment reversal into the global ground state, rather than higher-energy trapping.  The former assertion that all results regenerate from raw samples is superseded by the partial-reconstruction scope stated above.  The revision also corrects the classical-baseline scope and the interpretation of the finite-sample reference scale.  A worked pooling example and retrospective comparison of pooled, paired and per-preparation screening across both processors clarify the role of the observable pair, with a self-contained replay resource.  The campaign counts and initial-boundary distances are unchanged; the final-environment and screening comparisons are retrospective analyses.  Earlier protocol and calibration clarifications are retained.

\begin{acknowledgments}
This work received no external financial funding.  QPU access was provided through the D-Wave Leap developer platform; D-Wave had no role in study design, data analysis, interpretation, or manuscript preparation.  The author thanks the D-Wave support and training teams for platform access and technical assistance with the Leap environment.
\end{acknowledgments}

\subsection*{Conflicts of Interest}
The author declares no competing interests.

\subsection*{AI Use Disclosure}
AI tools assisted code scaffolding, language editing, computational reanalysis and manuscript auditing.  Documented tools across the development and revision history include OpenAI Codex (GPT-5.x historically and the GPT-6 family for the current revision), Anthropic Claude (Opus~4.x, Opus~5 and Fable~5.2), Moonshot Kimi (the configured K3 route), the historically configured Fugu Ultra route in September~2026, and Google Gemini~4 Neon.  The November editorial assessments used Gemini~4 Neon and Claude Fable~5.2.  Model-generated critiques informed checks and revisions; they are not evidence for the scientific claims.  The retrospective calculations and their verification scope are described in Methods and supplied with executable records.  The author is responsible for the scientific decisions and the accuracy, integrity and originality of the content.

\bibliography{refs}

\end{document}